%% file: C3PD_CLICdp.tex
\documentclass[a4paper,11pt]{article}
\pdfoutput=1 % if your are submitting a pdflatex (i.e. if you have
\usepackage{jinstpub} % for details on the use of the package, please
\usepackage{hyperref}
\usepackage{enumitem}
\usepackage{subcaption}
\usepackage{lineno}
\usepackage{floatrow}
\usepackage{siunitx}
\usepackage{graphicx}
\usepackage{caption,subcaption}
\usepackage{gensymb}

\usepackage{xcolor}
\usepackage{soul}

\title{Design and standalone characterisation of a capacitively coupled HV-CMOS sensor chip for the CLIC vertex detector}

\author[a,b,1]{I. Kremastiotis,\note{Corresponding author.}}
\author[a]{R. Ballabriga,}
\author[a]{M. Campbell,}
\author[a]{D. Dannheim,}
\author[a]{A. Fiergolski,} 
\author[a]{\newline D. Hynds,}
\author[a]{S. Kulis,}
\author[b]{I. Peric}

\affiliation[a]{CERN,\\Geneva, Switzerland}
\affiliation[b]{Karlsruhe Institute of Technology,\\Karlsruhe, Germany}

\emailAdd{iraklis.kremastiotis@cern.ch}

\abstract{The concept of capacitive coupling between sensors and readout chips is under study for the vertex detector at the proposed high-energy CLIC electron positron collider. The CLICpix Capacitively Coupled Pixel Detector (C3PD) is an active High-Voltage CMOS sensor, designed to be capacitively coupled to the CLICpix2 readout chip.
The chip is implemented in a commercial $180$~nm HV-CMOS process and contains a matrix of $128\times128$ square pixels with \SI{25}{\micro\meter} pitch. First prototypes have been produced with a standard resistivity of~$\sim20$~$\Omega$cm for the substrate and tested in standalone mode. The results show a rise time of $\sim20$~ns, charge gain of $190$~mV/ke$^{-}$ and $\sim40$~e$^{-}$ RMS noise for a power consumption of \SI{4.8}{\micro\watt}/pixel.
The main design aspects, as well as standalone measurement results, are presented.}

\keywords{ Analogue electronic circuits, Pixelated detectors and associated VLSI electronics,  Front-end electronics for detector readout, VLSI circuits }

\dedicated{This work was carried out in the framework of the CLICdp collaboration.}

\begin{document}

% generates the title page
\maketitle
\flushbottom

\input{./main_text.tex}

\input{./references.tex}
\end{document}

%% file: main_text.tex
\newcommand{\latex}{\LaTeX\xspace}

\section{Introduction}
\label{sec:Intro}

The Compact Linear Collider (CLIC) vertex detector requirements~\cite{reqs} are stringent, both in terms of performance and the challenging material budget, which is only~$0.2\%$~X$_0$ per detection layer. The material budget corresponding to the sensor and readout technology is only~$100$~\SI{}{\micro\meter} of silicon. 
Both the sensor and the readout ASICs have to be thinned down to about~$50$~\SI{}{\micro\meter} each in order to achieve this. In addition, as only air-flow cooling can be provided due to these material constraints, the power consumption of the system is limited to~$50$~mW/cm$^{2}$. This can be achieved by taking advantage of the low duty cycle of the CLIC beam, allowing the introduction of a pulsed powering scheme in order to keep the main driving stages of the front-end in a "power-off" state most of the time, from which they can be powered on quickly for the duration of the collisions.
Other requirements for the CLIC vertex detector include the spatial resolution of~$\sim3$~\SI{}{\micro\meter} and a time-stamp resolution of~$10$~ns. The pixel occupancy per bunch train is expected to reach up to $3\%$. 
%In the experiment, the electronics will be operated in room temperature.

Designing High Voltage (HV-) CMOS devices involves placing all of the on-pixel functionality inside a deep N-well, which both shields the electronics from the substrate (allowing the application of a substantial bias voltage) and acts as the charge collection node. This substrate bias allows the formation of a depletion region with a depth of several microns (depending on the wafer resistivity), providing fast charge collection. Charge collected on the N-well can be processed on-pixel by a Charge Sensitive Amplifier (CSA), and can then be digitised and read out either on the same chip (monolithic detector~\cite{peric}), or through coupling to a second readout chip~\cite{hvcmos}. Connecting the two chips capacitively has the advantage of avoiding the complex and costly solder bump-bonding process; the output signal of the HV-CMOS sensor is transferred to the input of the readout chip through a thin (a few~\SI{}{\micro\meter}) layer of glue applied between the two chips. On the other hand, this approach adds complications compared to the use of a passive sensor, as it requires more elaborate PCBs and wire-bonding schemes, due to the two ASICs operating simultaneously.

The HV-CMOS technology has been investigated in the context of the vertex detector studies for the proposed high-energy CLIC collider. A readout chip (CLICpix~\cite{clicpix}) has been designed, fabricated and tested extensively with a first-generation of capacitively coupled HV-CMOS sensors~\cite{ccpdv3}. 

In this paper, the design of a new sensor chip, the CLICpix Capacitively Coupled Pixel Detector (C3PD), is presented. The C3PD is designed to be compatible with the CLICpix2~\cite{clicpix2} readout chip, which is the successor of CLICpix and is based on the same front-end architecture as described in~\cite{clicpix}.
In order to match the footprint of the CLICpix2, C3PD contains a matrix of~${128\times128}$ pixels measuring~${25\times25}$~\SI{}{\micro\meter}$^{2}$, for a total matrix area of~${3.2\times3.2}$~mm$^{2}$. The chip was fabricated in a commercial~$180$ nm HV-CMOS process, using the standard substrate resistivity of ${\sim20}$~$\Omega$cm. This is a triple-well process, with all PMOS and NMOS transistors enclosed within the deep N-well.

\section{C3PD chip design}

\subsection{Design considerations}
\label{sec:Design}

\begin{figure}[htbp]
	\centering % \begin{center}/\end{center} takes some additional vertical space
	\includegraphics[width=.6\textwidth,trim=0 0 0 0,clip]{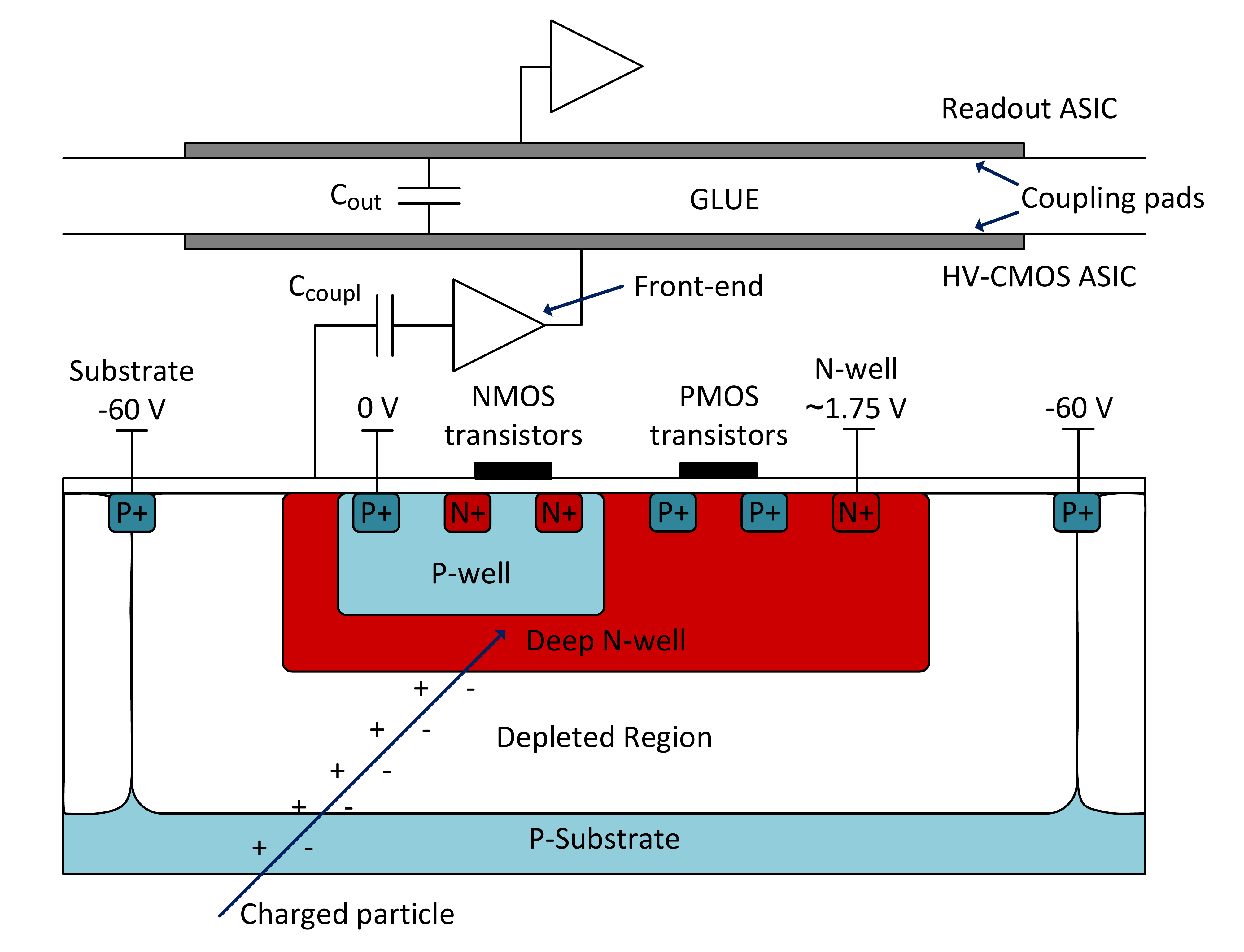}
	\caption{\label{fig:process} Schematic cross-section of a capacitively coupled assembly.}
\end{figure}

According to TCAD simulations~\cite{matthew}, a thickness of the order of~$\sim10$~\SI{}{\micro\meter} is expected for the depletion region with the standard substrate resistivity and a reverse bias of~$-60$~V. 
The input of the front-end is coupled to the deep N-well, so that the collected charge is integrated in the charge sensitive amplifier. The output of the amplifying circuit is routed to the coupling pad in order to be injected to the readout chip. A schematic cross-section of the assembly, including the wells where the NMOS and PMOS transistors are placed as well as the coupling between the sensor and readout chip, is shown in Figure~\ref{fig:process}.

A limitation of the technology used for the sensor circuit design is that any P+ diffusion in the N-well is coupled to the preamplifier input. The use of PMOS transistors is therefore minimised in order to avoid this parasitic coupling and to reduce undesired signal injection into the sensor. 

In order to suppress out-of-time background hits in the detector, a time slicing of hits with $10$~ns accuracy is required for the readout chip~\cite{reqs}.
To ensure that the output signal from the HV-CMOS sensor will be fully integrated within the integration time of the CLICpix2 readout chip ($\sim50$~ns), a fast charge injection is needed and therefore a rise time of order $20$~ns is aimed for.
%In order to meet the CLIC time-stamping requirement of $10$~ns and additionally to ensure that the output signal from the HV-CMOS is matched to the integration time of the CLICpix2 readout chip, a rise time of order $20$~ns is needed.

Simulations of the coupling capacitance between the two chips ($C_{out}$ in Figure~\ref{fig:process}) have shown that the expected capacitance is $\sim3$~fF, for a measured pad distance of~$3$~\SI{}{\micro\meter}. 
The space between the pads is mainly occupied by the passivation layer of the C3PD chip, with a thin layer of glue between the two chips~\cite{coupling}.

A most probable charge of $800$~e$^{-}$ was assumed to be deposited by a Minimum Ionising Particle (MIP) in the $10$~\SI{}{\micro\meter} thick depletion region, with collection via drift (without taking into account the charge collected via diffusion).
The charge gain of the amplifier should therefore be high enough to transfer a charge well above the minimum that can be detected by the readout chip, in order to compensate for charge sharing, straggling and to account for the expected noise levels of a few tens of electrons (simulated, as well as measured noise levels will be discussed in section~\ref{sec:tpulse}) and possible uncertainty in the values expected from simulations.
For a collected charge of $800$~e$^{-}$ and a charge gain of $120$~mV/ke$^{-}$, the pulse amplitude at the output of the sensor chip will be $96$~mV. Assuming a coupling capacitance of $3$~fF, this pulse will inject $\sim1.8$~ke$^{-}$ at the input of the readout chip, well above the simulated minimum threshold for the readout chip ($\sim600$~e$^{-}$). The detector is therefore expected to be fully efficient.

An overview of the requirements for the C3PD chip is presented in Table~\ref{tab:reqs}.

%The injected charge to the readout chip needs to be higher than the $600$~e$^{-}$ (or $\sim0.1$~fC) that correspond to the minimum threshold, as expected from simulations of the readout chip.

%Given simulations of the coupling capacitance between the two chips ($C_{out}$ in Figure~\ref{fig:process}) of $\sim3$~fF~\cite{coupling}, this would require a minimum output voltage pulse of $33$~mV. 

%In order to inject the minimum detectable charge to the readout chip for this most probable value and the simulated coupling capacitance of $\sim3$~fF a charge gain of $41$~mV/ke$^{-}$ would be necessary. 
%A safety factor of $3-5$ was added to this value in order to compensate for the charge sharing, the Landau fluctuations and to account for the expected noise levels of a few tens of electrons (simulated, as well as measured noise levels will be discussed in section~\ref{sec:tpulse}) and possible uncertainty in the values expected from simulations.

\begin{table}[htbp]
	\centering
	\caption{\label{tab:reqs} Requirements for the C3PD chip.}
	\smallskip
	\begin{tabular}{|l|r|}
		\hline
		  & Requirement\\
		\hline
		Pixel size 			& $25\times25\ $ \SI{}{\micro\meter}$^{2}$	\\
		Sensor thickness 	& $50\ $ \SI{}{\micro\meter} (after thinning)\\
		Rise time 			& $\sim20\ $ ns			\\
		Charge gain 		& $>120\ $ mV/ke$^{-}$		\\
		Power pulsing		& Required \\
		\hline
	\end{tabular}
\end{table}

\subsection{Pixel design}
\label{sec:Pixel}

\begin{figure}[t!]
	\centering % \begin{center}/\end{center} takes some additional vertical space
	\includegraphics[width=.5\textwidth,trim=80 35 100 520,clip]{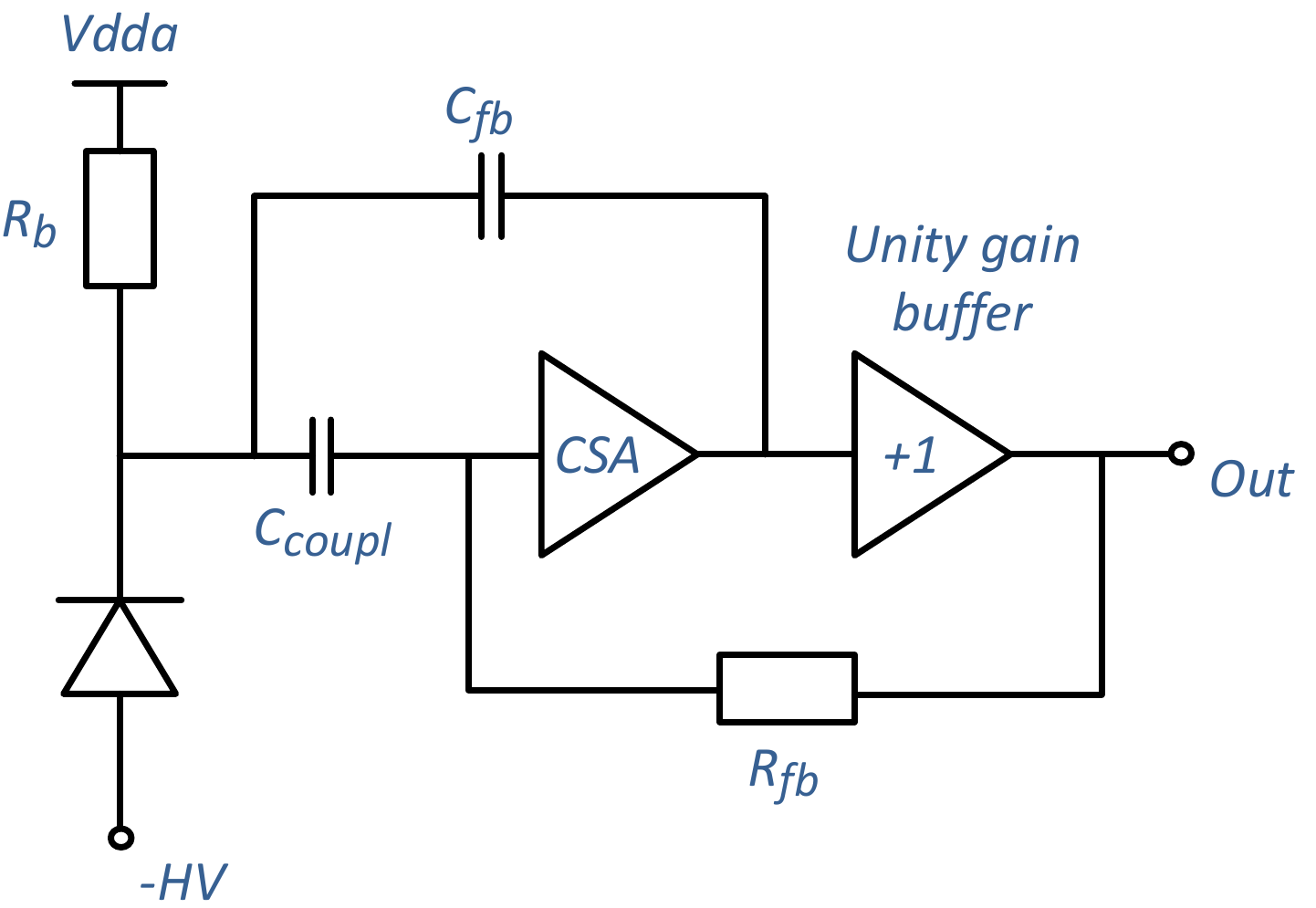}
	\caption{\label{fig:block} C3PD pixel block diagram.}
\end{figure}

\begin{figure}[b!]
	\centering % \begin{center}/\end{center} takes some additional vertical space
	\includegraphics[width=.7\textwidth]{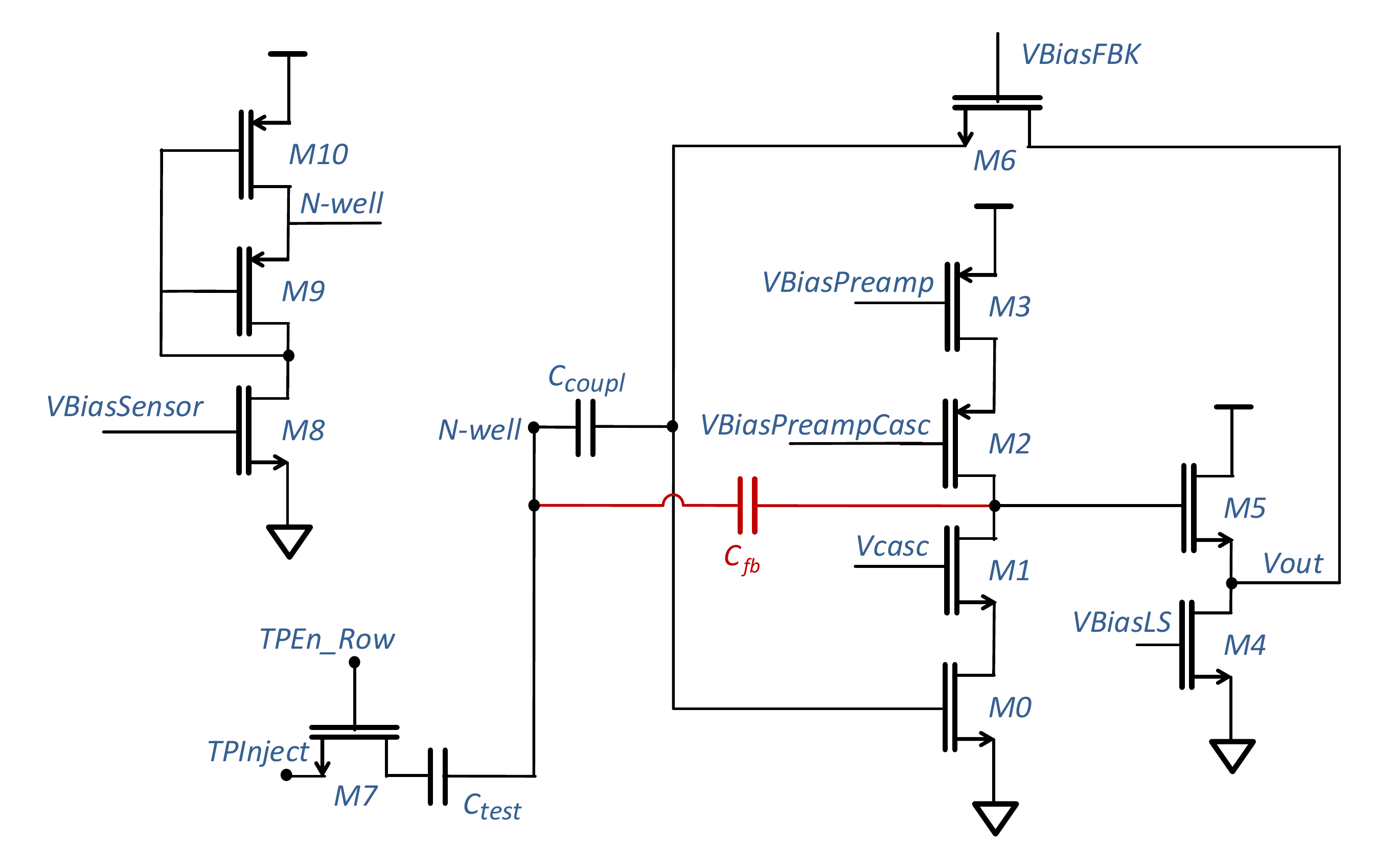}
	\caption{\label{fig:schematic} C3PD pixel schematic. The parasitic feedback capacitance is shown in red colour.}
\end{figure}

The C3PD pixel was designed following the specifications listed above. 
A block diagram of the pixel is shown in Figure~\ref{fig:block} and the detailed pixel schematic is illustrated in Figure~\ref{fig:schematic}.
The charge deposited in the diode is amplified by a charge sensitive amplifier (CSA) (transistors M0$-$M3 in Figure~\ref{fig:schematic}), which is followed by a unity gain buffer (transistors M4, M5) placed inside the feedback loop in order to maintain a fast rise time after loading the circuit with a capacitance, and to improve the pixel-to-pixel homogeneity. The phase margin for this design is $>75\degree$ (as simulated for different corners). A single power supply is used for the front-end, where an NMOS device (M0 in Figure~\ref{fig:schematic}) is used as input transistor. An NMOS transistor was selected for the input as it is expected to have better performance in terms of power supply rejection ratio.
The sensor capacitance for such a pixel size in this process is expected to be up to $40$~fF, taken as a conservative (worst case) value.
 
A power pulsing scheme has been introduced, where the most power consuming nodes of the analogue front-end (CSA and unity gain buffer) can be set to a "power-off" state between subsequent bunch trains. This is realised by means of multiplexing "power-on" and "power-off" biasing DACs for each node, in order to switch between the two states.
The possibility to inject test pulses with a programmable amplitude to individual pixels was also implemented for the C3PD chip. 

\begin{figure}[htbp]
	\centering % \begin{center}/\end{center} takes some additional vertical space
	\includegraphics[width=.5\textwidth]{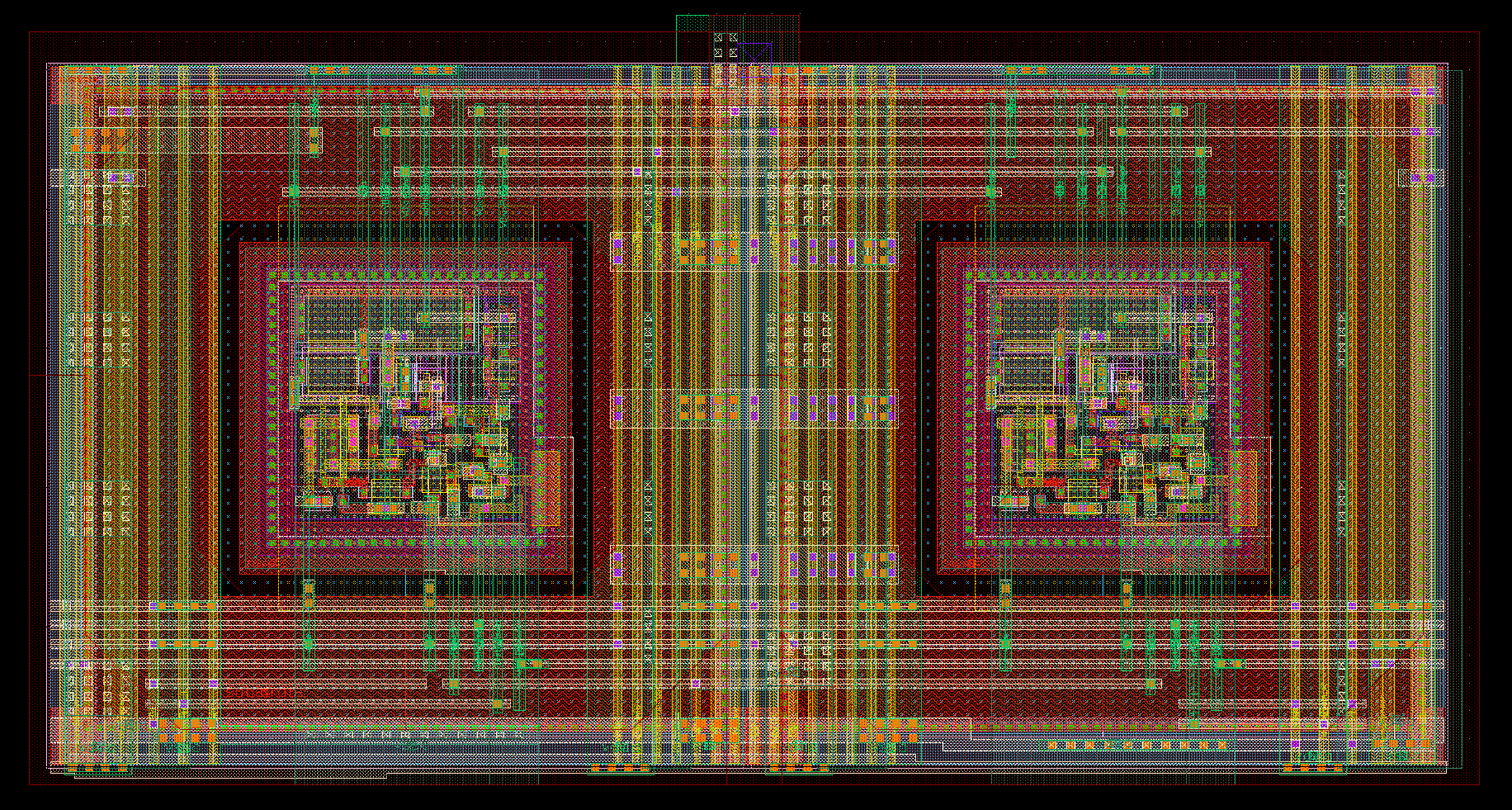}
	\caption{\label{fig:layout} C3PD double pixel layout. The pixels are shown to the left and the right side. The biasing lines are routed between the pixels.}
\end{figure}

As explained in section~\ref{sec:Design}, the number of PMOS transistors has been kept as low as possible. Aside from the ones in the sensor biasing circuit (transistors M9, M10 in Figure~\ref{fig:schematic}), only 2 PMOS transistors (M2, M3) are used, which have their common diffusion merged in the layout, thus minimising the circuit area covered by P+ diffusion. The parasitic feedback capacitance ($C_{fb}$, shown in red in Figure~\ref{fig:schematic}) is the drain to N-well capacitance of M2.
The transistor M6 serves as the resistive feedback ($R_{fb}$ in figure~\ref{fig:block}), providing a continuous reset for the circuit.
Care was put into optimising the pixel layout, since such a design is sensitive to parasitic capacitances and different ways of placing the devices and routing the signals would likely result in significant performance variations. The pixel was therefore simulated systematically using the extracted layout view, in order to take into account the parasitic capacitances. In this way the layout was optimised by means of minimising the parasitic coupling to the sensor. 
The pixels are organised in a double column structure where adjacent pixels share common biasing lines. The layout of a double pixel is shown in Figure~\ref{fig:layout}.

\subsection{Pixel matrix layout}

The C3PD matrix consists of $128\times128$ pixels, arranged in 64 double columns, with $25$~\SI{}{\micro\meter} pitch. Three slightly different variations of the pixel can be found within the matrix for prototyping purposes. The main part of the matrix ($62$ double columns) consists of regular pixels, while the two rightmost double columns include structures for testing different flavours of pixels. 
The first of the testing columns comprises pixels with a metal-to-metal coupling capacitance ($C_{coupl}$ in Figure~\ref{fig:schematic}) instead of a CMOS gate capacitance. The expected benefits of the metal-to-metal capacitor are its linearity and the fact that the circuit can have a larger coupling capacitance while making more area available for active devices.
The second testing column features a modified sensor biasing scheme, with a single PMOS transistor used to bias the sensor in order to simplify the design. 
The performance of these test structures will be studied using capacitively coupled assemblies with the readout chip.

A cluster of $3\times3$ monitoring pixels is placed at the bottom-right edge of the 62 regular double columns, as will be described in section~\ref{sec:interface}. The C3PD matrix also features alignment marks, matching those on the readout chip. This will allow precise alignment of the two chips during flip-chip assembly.

\begin{figure}[htbp]
	\centering % \begin{center}/\end{center} takes some additional vertical space
	\includegraphics[width=.3\textwidth,trim=135 30 70 270,clip]{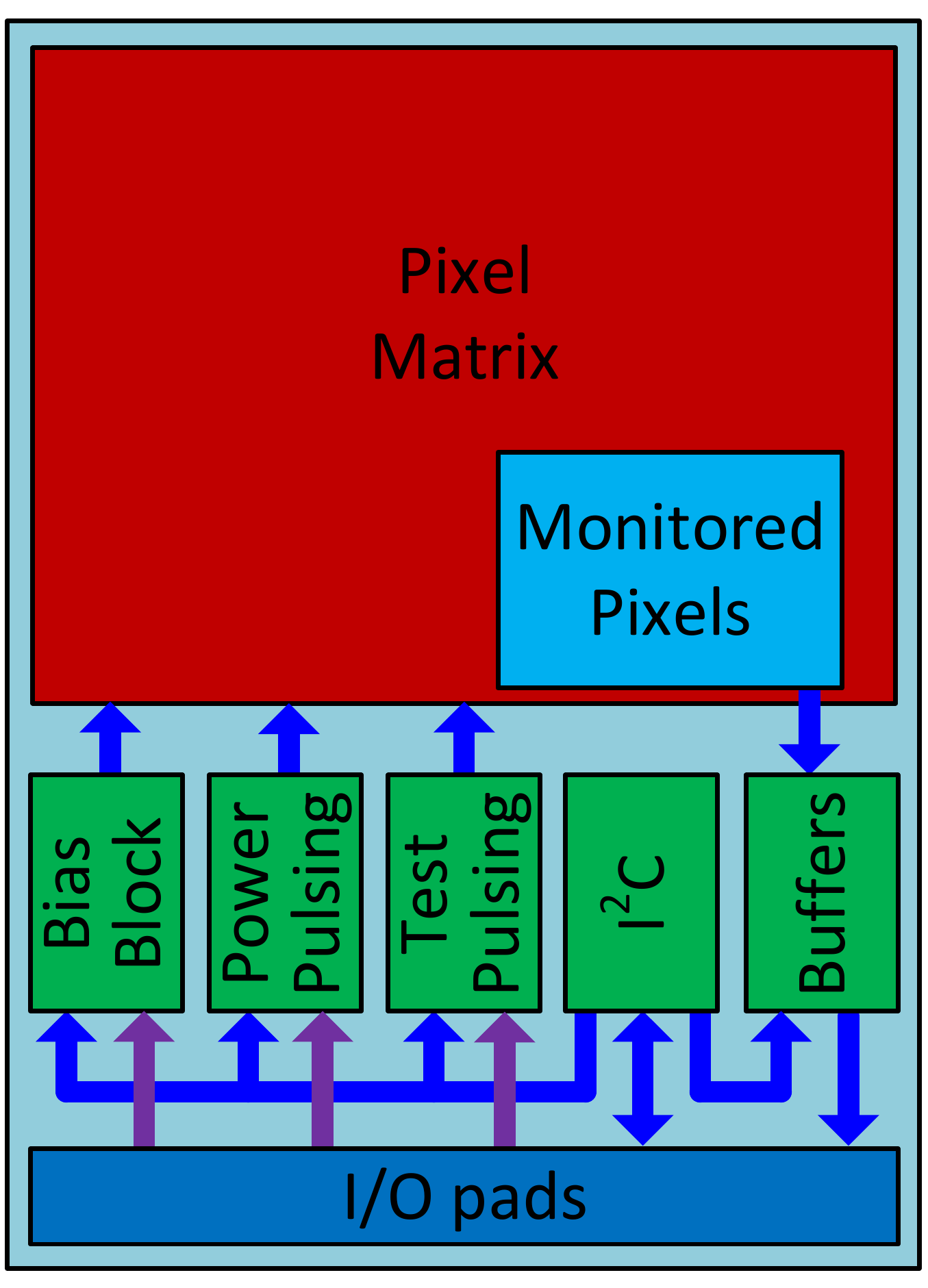}
	\caption{\label{fig:chip} C3PD chip block diagram.}
\end{figure}

\subsection{Chip interface}
\label{sec:interface} 

A block diagram of the chip is presented in Figure~\ref{fig:chip}. In the digital domain, a standard I$^2$C interface has been implemented which is responsible for controlling the biasing DACs, the power pulsing and the test pulse generation~\cite{digital,i2c}. 
Two address pins are used in order to define different chip addresses on the I$^2$C bus, opening the possibility of multi-chip modules. In the analogue domain, the possibility to monitor voltages produced by the on-chip DACs has been implemented. Similarly, any of the voltages produced by on-chip DACs can be overwritten by an external source.

In order to study the analogue performance in standalone mode, a $3\times3$ cluster of pixels with direct readout of the preamplifier output has been implemented, with the outputs multiplexed in order to read them out in different configurations. Depending on the chosen configuration, the outputs of four pixels from this cluster are monitored simultaneously in different patterns. Unity gain buffers have been placed in the analogue periphery in order to drive these signals to the wire-bond pads. 
A dedicated pixel has also been implemented where the test pulse voltage injected into the pixel (that would typically reach the test pulse injection capacitance, $C_{test}$, in Figure~\ref{fig:schematic}) is directly connected to the coupling pad. This can be used to directly measure the capacitance between the C3PD and the readout chip, using the known voltage on this pad and measuring the calibrated charge deposited on the readout ASIC. 

%The pad providing high-voltage biasing for the substrate is located at the edge of the I/O array. This pad is separated from all other pads, in order to avoid electrostatic discharges (ESD).

\section{C3PD characterisation}

The C3PD chip was tested using a custom designed setup, comprising a PCB to which the chip was wire-bonded, an interface board providing all necessary control signals and power supplies~\cite{uasic}, and an FPGA development board. 

\begin{figure}[htbp]
	\centering % \begin{center}/\end{center} takes some additional vertical space
	\includegraphics[width=.4\textwidth,trim=0 0 0 0,clip]{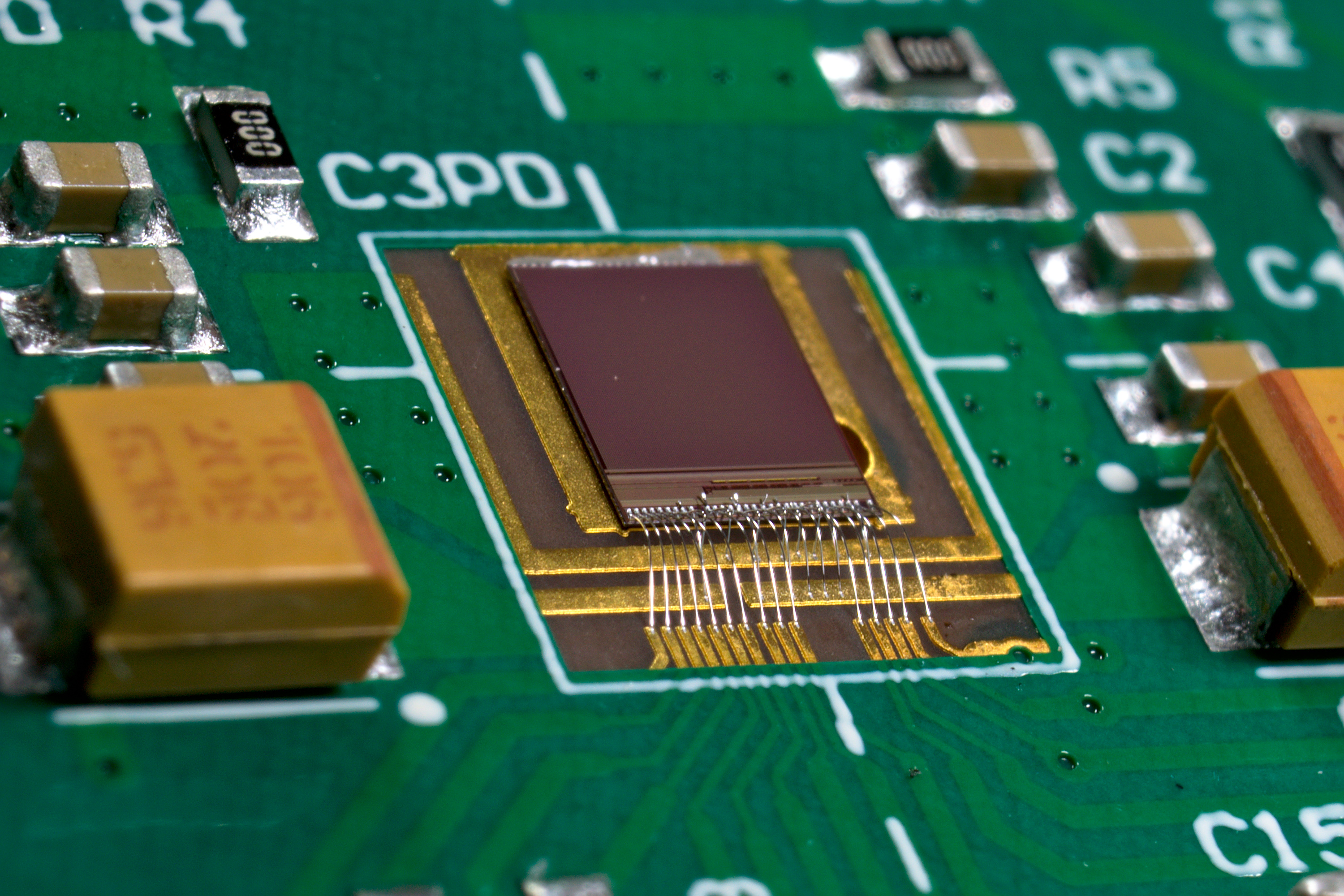}
	\caption{\label{fig:board} C3PD chip thinned down to $50$~\SI{}{\micro\meter}, mounted and wire-bonded on the PCB.}
\end{figure}

Testing of the C3PD front-end was restricted to the monitored cluster of pixels in standalone mode, without bonding the sensor to a readout chip. 
The characterisation of the C3PD chip was performed using test pulses, as well as a $^{55}$Fe source. In addition to the standard thickness ($250$~\SI{}{\micro\meter}), samples thinned down to $50$~\SI{}{\micro\meter} have been received and tested. A $50$~\SI{}{\micro\meter} thin chip wire-bonded on the PCB is shown in Figure~\ref{fig:board}. 
Measurement results with the C3PD chip in standalone mode are presented in the following sections. A continuous power (without power pulsing) was used for this set of measurements, unless otherwise stated.

\subsection{I-V characteristics}

\begin{figure}[htbp]
	\centering % \begin{center}/\end{center} takes some additional vertical space
	\includegraphics[width=.6\textwidth,trim=0 0 0 0,clip]{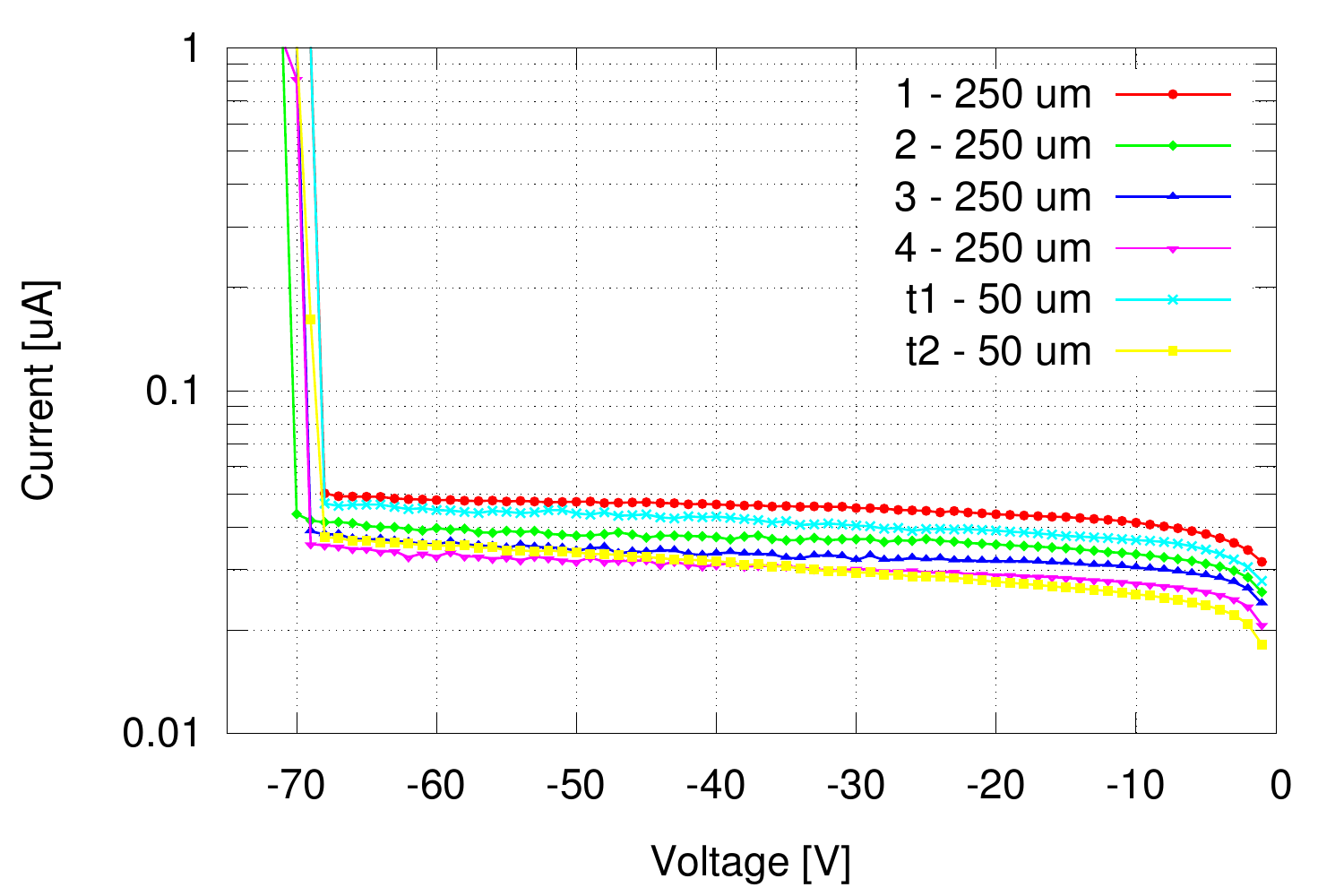}
	\caption{\label{fig:iv} I-V curves of the measured C3PD chips.}
\end{figure}

The applied sensor bias voltage was scanned in order to study the I-V characteristics of the sensor.
Figure~\ref{fig:iv} presents the measured leakage current as a function of the reverse bias, for four standard thickness samples and two samples thinned to $50$~\SI{}{\micro\meter}.
The mean value of the sensor leakage current at the nominal sensor bias of $-60$~V is $\sim40$~\SI{}{\nano\ampere}, measured for the full chip (corresponding to less than $2.5$~\SI{}{\pico\ampere} per pixel) at room temperature (which is also the foreseen operating temperature in the experiment). The corresponding power consumption is negligible compared to the total power consumed by the chip.
The breakdown occurred at a voltage between $-68$ and $-70$~V for all measured assemblies. 
The measured values for the leakage current and the breakdown voltage are suitable for operating the chip at the nominal high voltage bias of $-60$~V.

\subsection{Measurements with test pulses}
\label{sec:tpulse}

\begin{figure}[htbp]
	\centering
	\begin{subfigure}[b]{.49\textwidth}
		\centering % \begin{center}/\end{center} takes some additional vertical space
		\includegraphics[width=1\textwidth]{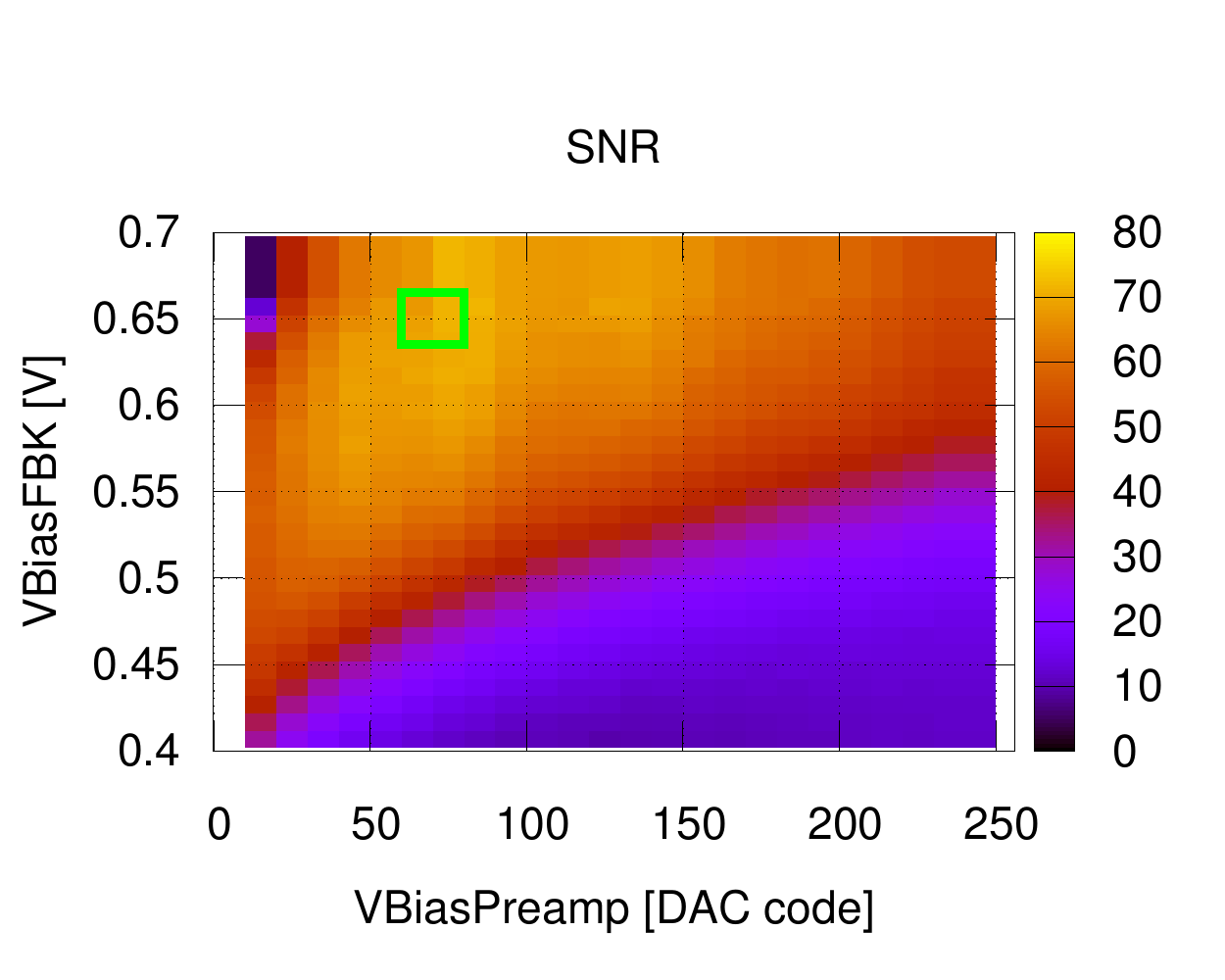}
		\caption{\label{fig:scana}}
	\end{subfigure}
	\begin{subfigure}[b]{.49\textwidth}
		\centering % \begin{center}/\end{center} takes some additional vertical space	
		\includegraphics[width=1\textwidth]{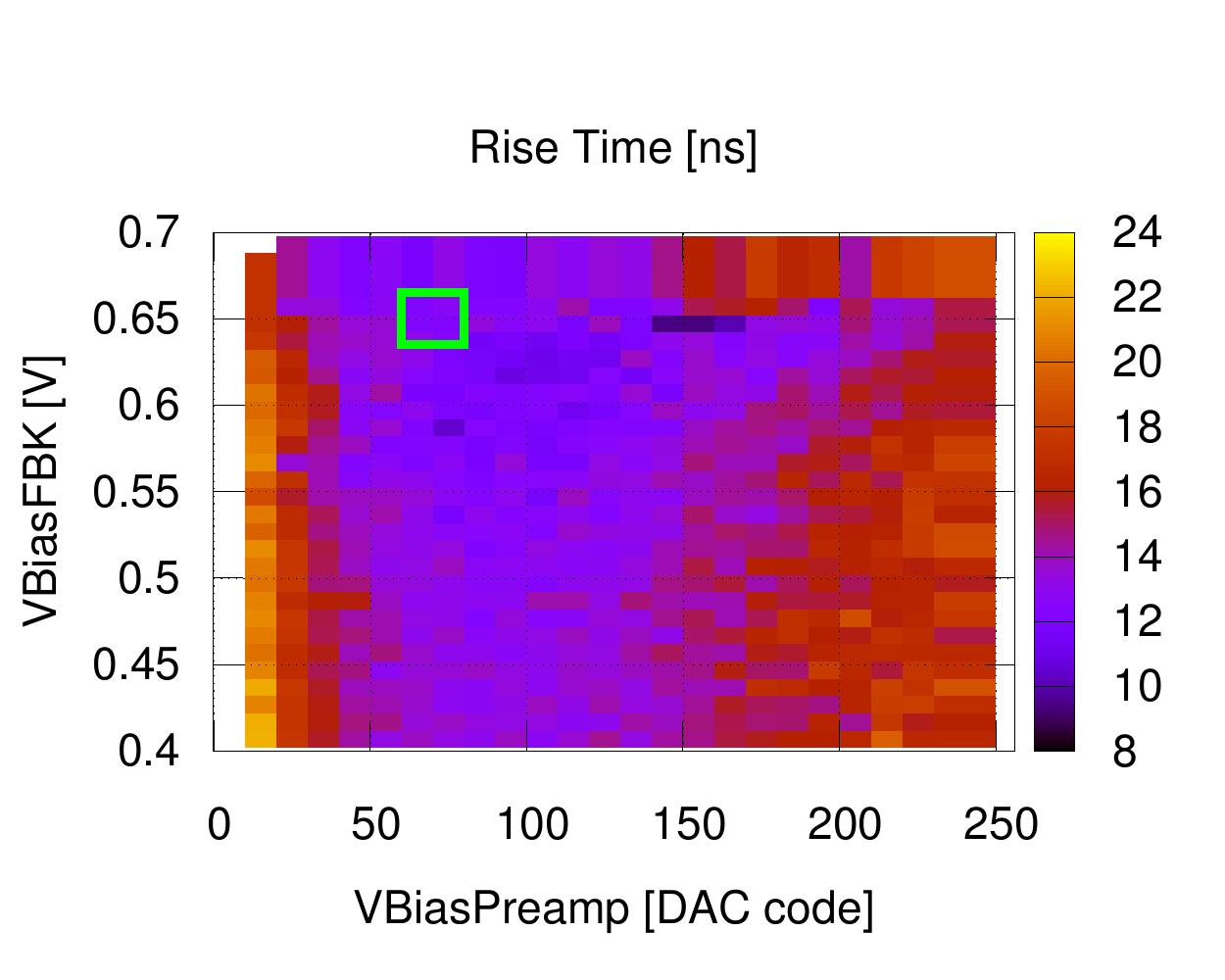}
		\caption{\label{fig:scanb}}
	\end{subfigure}
	\caption{\label{fig:scan}~(\subref{fig:scana}) Signal-to-noise ratio and~(\subref{fig:scanb}) rise time as measured for different biasing points of the preamplifier and the feedback. The chosen operating point is marked in green.} 
\end{figure}

Test pulses were injected in order to characterise the analogue performance for different operating points of the front-end. The operating point was configured scanning the on-chip DACs that are used to bias the different nodes of the front-end (in particular the preamplifier, the unity gain buffer and the feedback transistor). The results of these scans were used in order to determine the optimum biasing settings with respect to the main performance parameters of the chip (amplitude, noise, rise time of the pixel output pulse and power consumption). 
Test pulses were injected to each pixel individually during these calibration measurements. 
Two characteristic plots are shown in Figure~\ref{fig:scan}, where the signal-to-noise ratio and rise time are measured for different biasing points of the preamplifier and the feedback transistor.
The code of the preamplifier biasing DAC is shown on the x-axis. On the y-axis, the feedback biasing was scanned providing an external voltage (overwriting the internal DAC) in order to achieve a higher range for the voltage provided to the feedback transistor.

\begin{figure}[htbp]
	\centering % \begin{center}/\end{center} takes some additional vertical space
	\includegraphics[width=.55\textwidth,trim=0 0 0 0,clip]{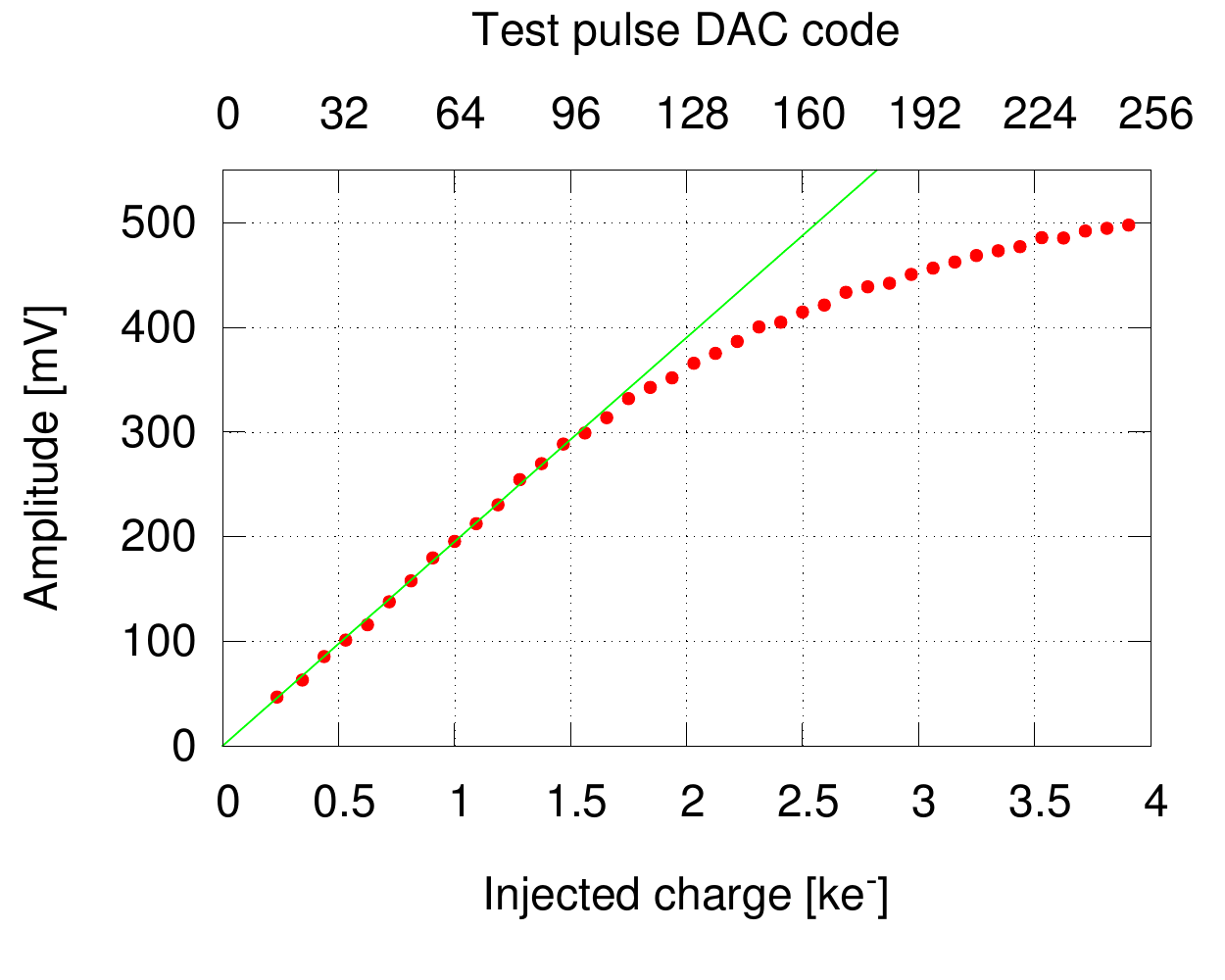}
	\caption{\label{fig:tp} Output amplitude, averaged over 64 pulses, as a function of the injected test pulse DAC code and corresponding injected charge.}
\end{figure}

Figure~\ref{fig:tp} presents the measured mean amplitude of 64 samples at the pixel output as a function of the injected charge. The corresponding DAC code is shown on the second x-axis. The injected charge was extracted knowing the test capacitance (see section~\ref{sec:capacitance}) and monitoring the voltage level of the test pulse as a function of the DAC code. As indicated in Figure~\ref{fig:tp}, the dependence of the pixel output amplitude on the injected input charge is linear for an injected charge up to about $1.5$~ke$^{-}$. This dependence was simulated to be linear for an injected charge up to $2$~ke$^{-}$. The difference in linearity can be explained due to the difference between the measured and simulated charge gain (which is consistent with the disagreement in the feedback capacitance, as described in section~\ref{sec:capacitance}).

\begin{figure}[htbp]
	\centering % \begin{center}/\end{center} takes some additional vertical space
	\includegraphics[width=.55\textwidth,trim=0 0 0 0,clip]{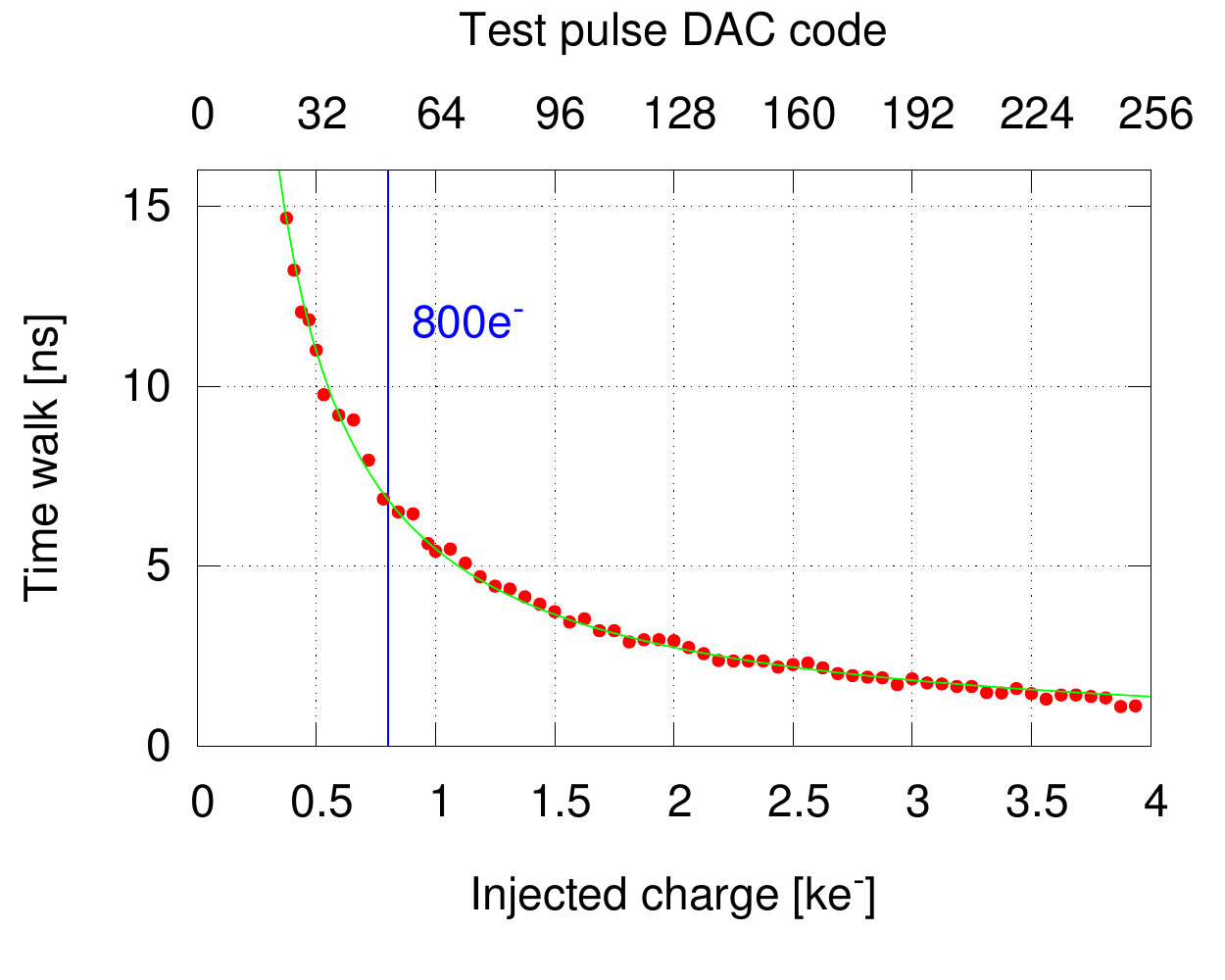}
	\caption{\label{fig:tw} Time walk, averaged over 64 pulses, as a function of the injected test pulse DAC code and corresponding injected charge.}
\end{figure}

In Figure~\ref{fig:tw} the time walk is plotted as a function of the injected charge. For this measurement, the trigger threshold of the oscilloscope was set to $33$~mV, which corresponds to the amplitude which will inject the minimum detectable charge to the readout chip as explained in section~\ref{sec:Design}. The time was measured between the moment when the test pulse strobe was sent and when the output signal crossed the threshold value. The vertical line at $800$~e$^{-}$ represents the expected average charge deposited by a MIP in the depletion region. In this plot, the offset resulting from the delay of the cables has been subtracted, in order to have a better view on the threshold crossing time depending on the input charge.
As shown in Figure~\ref{fig:tw}, the time walk for charges above $500$~e$^{-}$ is within the $10$~ns time-stamping requirement. For charges below that value time walk correction can be applied using the energy information (Time-over-Threshold) provided by the readout chip. 

For the above settings an average current consumption of $2.67$~\SI{}{\micro\ampere} per pixel was measured, of which $2.44$~\SI{}{\micro\ampere} is consumed by the preamplifier, and $0.23$~\SI{}{\micro\ampere} by the unity gain buffer. The nominal operating voltage for the C3PD chip is $1.8$ V. 
Introducing the power pulsing scheme can set the main driving nodes of the analogue front-end to a "power-off" mode, resulting in an average current consumption of only $53$~\SI{}{\nano\ampere} per pixel.
A delay of about $15$~\SI{}{\micro\second} after the power enable signal has been measured (by injecting test pulses with a controlled delay after the power enable signal) to be sufficient for the circuit to be ready to detect particles, without any noticeable effect on noise.
Therefore, assuming a duty cycle length of $30$~\SI{}{\micro\second} (following a conservative approach and providing a duty cycle twice as long as the measured power-up time) over the $20$~ms between subsequent bunch trains of the CLIC accelerator, the average power consumption for the matrix over the $50$~Hz cycle will be $\sim16$~mW/cm$^{2}$. The average power consumption during the experiment conditions is dominated by the "power-off" state, and could be minimised in future versions by optimising the range of the power-off DACs.
A reasonably low power consumption during the "power-on" state is required in order to match the instantaneous power consumption that can be delivered to the detector, which is foreseen to reach a few W/cm$^{2}$~\cite{power},
and also to have the possibility to operate the chip in lab and beam tests with continuous power and without the use of additional cooling. Additionally, in view of future HV-CMOS developments, the power consumption can be used as a reference for applications with different powering schemes.
The average power consumption of the readout chip is expected to be at a comparable level.

The results shown in Table~\ref{tab:meas} present the average values and standard deviations of the main pixel characteristics for all monitored pixels from four standard thickness and two thinned down assemblies. 
A test pulse of $1.63$ ke$^{-}$ (charge equivalent to the most probable energy of the photons from an $^{55}$Fe source, as will be explained in section~\ref{sec:capacitance}) was injected for the amplitude and rise time measurements.
No systematic difference has been observed between standard and thinned samples. 

Simulations of the front-end were performed under the same conditions and the simulated values are presented in the rightmost column of Table~\ref{tab:meas}. 
Regarding the mismatch between the simulated and the measured rise time, it can be attributed to a convolution of different factors, from which the dominant one is believed to be additional parasitic capacitances loading the monitored outputs. The buffers which are driving the signal to the readout have been simulated to be fast enough, and are therefore expected to have negligible contribution to the degradation of rise time. The measured rise time is, according to Monte Carlo simulations, close to the three-sigma value.
Measured, as well as simulated values presented in Table~\ref{tab:meas} refer to the monitored pixels. For the regular pixels in the matrix the rise time was simulated to be $\sim30\%$ faster.  
The feedback capacitance depends on the drain voltage of transistor M2, which is set by the the $V_{gs}$ of M4, the $V_{ds}$ of M6 and the $V_{gs}$ of M0 in Figure~\ref{fig:schematic}. Uncertainties in these values can result to an extracted $C_{fb}$ different than the one present in the circuit. 
As the charge gain is inversely proportional to the feedback capacitance~\cite{wermes}, a difference in $C_{fb}$ can explain the discrepancy between the simulated and the measured charge gain.

\begin{table}[t!]
	\centering
	\caption{\label{tab:meas} Pixel characteristics for monitored pixels, averaged over 6 assemblies, measured with test pulses of $1.63$ ke$^{-}$ charge. The simulated values are shown in the rightmost column.}
	\smallskip
	\begin{tabular}{|l|r|r|r|}
		\hline
		Parameter & Average & Standard deviation & Simulated \\
		\hline
		Amplitude [mV] 									& $302$		& $30.9$	& $210$		\\
		Noise RMS [e$^{-}$] 							& $40$		& $3.8$		& $45$		\\
		Rise time [ns]									& $20.8$	& $3.75$	& $13.5$	\\
		Power/pixel, on state [\SI{}{\micro\watt}]		& $4.8$		& $0.41$	& $4.6$		\\
		Power/pixel, off state [\SI{}{\micro\watt}]		& $0.095$	& $0.03$	& $0.107$	\\
		\hline
	\end{tabular}
\end{table}

\subsection{Measurements with source}

\begin{figure}[htbp]
	\centering % \begin{center}/\end{center} takes some additional vertical space
	\begin{subfigure}{1\textwidth}
		\centering % \begin{center}/\end{center} takes some additional vertical space
		\includegraphics[width=.63\textwidth]{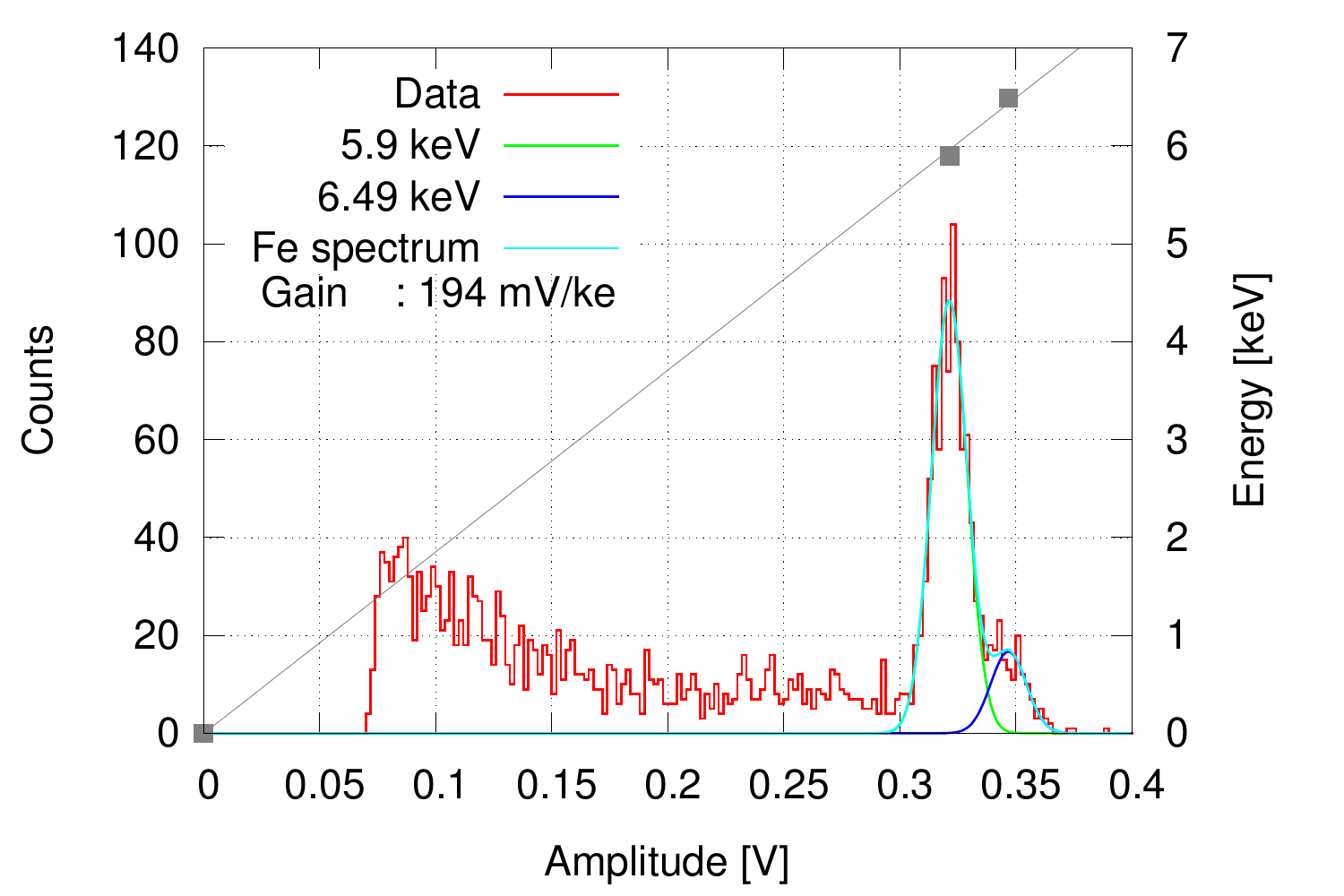}
		\subcaption{\label{fig:fea}}
	\end{subfigure}

	%	\qquad
	\begin{subfigure}{1\textwidth}
		\centering % \begin{center}/\end{center} takes some additional vertical space	
		\includegraphics[width=.58\textwidth]{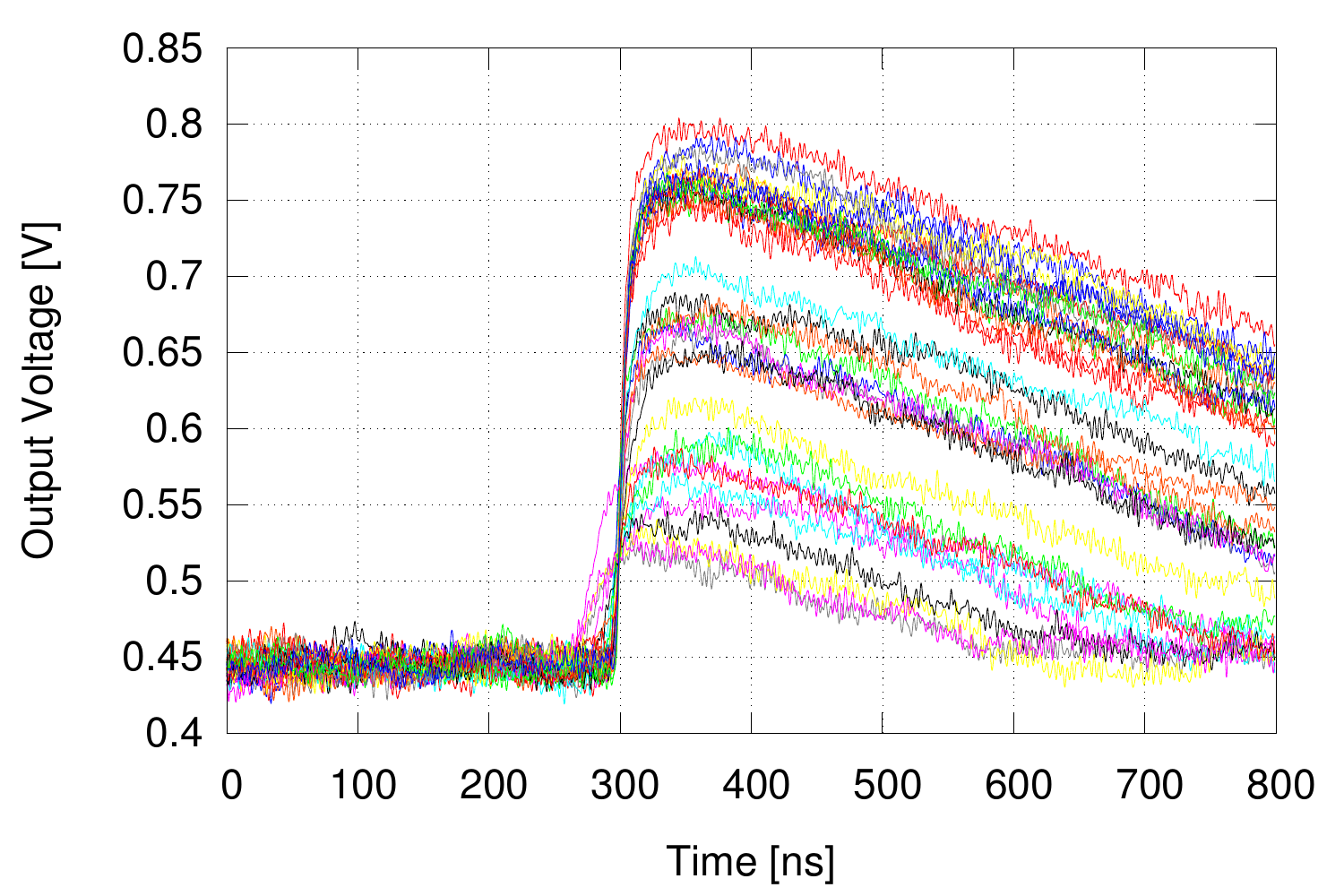}
		% "\includegraphics" from the "graphicx" permits to crop (trim+clip)
		% and rotate (angle) and image (and much more)
		\subcaption{\label{fig:feb} }
	\end{subfigure}
	\caption{\label{fig:fe}~(\subref{fig:fea}) Resulting amplitude spectrum and double Gaussian fit for one of the monitored pixels and~(\subref{fig:feb}) sample pulses from a $^{55}$Fe source.} 
\end{figure}

In Figure~\ref{fig:fe}, the pulse amplitude spectrum from a $^{55}$Fe source is presented, measured with a $20$~$GSa/s$ sampling oscilloscope on one of the monitored pixels, along with a sample of the output pulse shapes. 
The sum of two Gaussian distributions is fitted to the sampled amplitudes.
As shown, the two peaks of the fit correspond to the two most probable energies deposited by the photons from the $^{55}$Fe source ($5.9$~keV and $6.49$~keV), resulting in a gain of $\sim190$~mV/ke$^{-}$.
A tail due to charge sharing can be seen in the measured spectrum, which is the result of measurements from only a single monitored pixel.

\subsection{Feedback and test capacitance calculation}
\label{sec:capacitance} 

The results from the $^{55}$Fe source measurement were used to estimate the feedback capacitance of the C3PD amplifier. With a known $^{55}$Fe deposited energy of $5.9$~keV and electron-hole pair creation energy in silicon of $3.62$~eV, we can expect a deposited charge in the detecting volume equal to $0.26$~fC (or $1.63$~ke$^{-}$). The voltage peak corresponding to this energy is at $\sim320$~mV, as taken from the mean value of the first Gaussian in Figure~\ref{fig:fe}, which gives a charge gain of  $\sim190$~mV/ke$^{-}$. From the voltage peak, we can estimate a value for the feedback capacitance:

\begin{equation}
\label{eq:cfb}
C_{fb} =\dfrac{0.26fC}{320mV}\approx 0.81 fF \,
\end{equation}

In order to estimate the test pulse injection capacitance, $C_{test}$, a scan of the test pulse DAC code was performed while monitoring the pixel output (Figure~\ref{fig:tp}). From this, the slope of the output voltage amplitude as a function of the test pulse DAC code was extracted. At the DAC code for which the measured output voltage is close to the $320$~mV corresponding to the first peak of the $^{55}$Fe source spectrum, one can assume that a charge of $1.63$~ke$^{-}$ is injected at the input of the preamplifier. Once this DAC code was measured, it was possible to monitor the dedicated pixel where the injected test pulse is output (see section~\ref{sec:interface}) to determine the amplitude of the injected test pulse. It was observed that an injected test pulse of $370$~mV results in an output amplitude close to the one measured during tests with the $^{55}$Fe source. It is therefore a good approximation that a $370$~mV test pulse will inject $1.63$~ke$^{-}$, resulting in an injection capacitance of $\sim0.70$~fF.

The above procedure was followed for several other samples in order to characterise their performance with the $^{55}$Fe source and estimate their feedback and test capacitances. The results give average values of $C_{fb}=0.83$~fF~$\pm15\%$ and $C_{test}=0.70$~fF~$\pm2\%$, which are in good agreement with the simulated post-layout extracted values ($1.2$~fF for $C_{fb}$ and $0.8$~fF for $C_{test}$). 
The analysis followed for estimating the $C_{fb}$ assumes an infinite open loop gain and no continuous reset. As calculated, this assumption leads to a feedback capacitance overestimated by $\sim7.5\%$, mainly due to the continuous reset that is present in the circuit (provided by transistor M6, as described in section~\ref{sec:Pixel}).

\section{Summary and future steps}
The C3PD chip is a next-generation HV-CMOS sensor for the CLIC vertex detector R\&D. Results in standalone tests have shown a rise time of the pixel output pulse of the order of $20$~ns.
The average charge gain was measured to be $190$~mV/ke$^{-}$ with an RMS noise of $40$~e$^{-}$. 
These results match the detector requirements. 
The power consumption per pixel is $4.8$~\SI{}{\micro\watt} and, taking into account the introduced power pulsing scheme, the average power consumption for the matrix over the $50$~Hz cycle is $\sim16.4$~mW/cm$^{2}$.
The resulting average power consumption provides enough margin for the readout chip in view of the target of $50$~mW/cm$^{2}$ total average power consumption of the capacitively coupled assemblies. 
%A first estimation for the readout chip gives, for the same power pulsing scheme, an average power consumption of $\simXX$~mW/cm$^{2}$.

Samples of the C3PD chip have also been thinned to $50$~\SI{}{\micro\meter}, and have been successfully tested along with the standard $250$~\SI{}{\micro\meter} thick chips with no observed changes in their performance. All assemblies have so far shown only small variations between devices in terms of rise time, power consumption, signal-to-noise ratio and leakage current.

The next steps for C3PD will be to produce new versions of the chip with higher substrate resistivity, in order to study the expected beneficial effects on the sensor performance. Further tests, such as a study on gain and noise variations over the matrix, will take place using assemblies with the CLICpix2 readout chip. Test beam characterisation of the capacitively coupled assemblies will be performed in the coming months.

\section*{Acknowledgements}
This project has received funding from the European Union's Horizon 2020 research and innovation programme under grant agreement No 654168.